\pdfoutput=1
\documentclass[lettersize,journal]{IEEEtran}

\IEEEoverridecommandlockouts
\usepackage{amsfonts}
\usepackage[dvips]{graphicx}
\usepackage{times}
\usepackage{cite}
\usepackage{amsmath}
\usepackage{cases}
\usepackage{array}
\usepackage{dsfont}
\usepackage{amssymb}

\usepackage[normalem]{ulem}
\usepackage{stfloats}
\usepackage{graphicx}
\usepackage{subfigure}
\usepackage{pdfpages}
\usepackage{footnote}
\usepackage{booktabs}
\usepackage{array}
\usepackage{multirow}
\usepackage{bm}
\usepackage{empheq}
\usepackage{amsthm}
\usepackage{algorithm}
\usepackage{algorithmic}
\usepackage{color}
\usepackage{diagbox}
\usepackage{caption}
\theoremstyle{plain}

\theoremstyle{definition}

\usepackage{cite}
\usepackage{amsmath,amssymb,amsfonts}
\usepackage{algorithmic}
\usepackage{graphicx}
\usepackage{textcomp}
\usepackage{xcolor}

\usepackage{makecell}

\def\BibTeX{{\rm B\kern-.05em{\sc i\kern-.025em b}\kern-.08em
    T\kern-.1667em\lower.7ex\hbox{E}\kern-.125emX}}
\usepackage{balance}

\ifCLASSINFOpdf
\else
\fi

\begin{document}

\title{ SNR-EQ-JSCC: Joint Source-Channel Coding with SNR-Based Embedding and Query }







\author{\IEEEauthorblockN{$\text{Hongwei Zhang}$,~\IEEEmembership{Graduate Student Member,~IEEE}, $\text{Meixia Tao}$,~\IEEEmembership{Fellow,~IEEE}}
\vspace{-25pt}

\thanks{Corresponding author: Meixia Tao.}

\thanks{The authors are with the Department of Electronic Engineering and the Cooperative Medianet Innovation Center (CMIC), 
Shanghai Jiao Tong University, Shanghai, China (e-mails: \{zhanghwei, mxtao\}@sjtu.edu.cn).}

\thanks{This work is supported by the National Natural Science Foundation of China under grant 62125108 and by the Fundamental Research Funds for the Central Universities of China.}
}

\maketitle

\begin{abstract}

Coping with the impact of dynamic channels is a critical issue in joint source-channel coding (JSCC)-based semantic communication systems. In this paper, we propose a \textcolor{black}{lightweight} channel-adaptive semantic coding architecture called SNR-EQ-JSCC. It is built upon the generic Transformer model and achieves channel adaptation (CA) by Embedding the signal-to-noise ratio (SNR) into the attention blocks and dynamically adjusting attention scores through channel-adaptive Queries. Meanwhile, penalty terms are introduced in the loss function to stabilize the training process. \textcolor{black}{Considering that instantaneous SNR feedback may be imperfect, we propose an alternative method that uses only the average SNR, which requires no retraining of SNR-EQ-JSCC.} Simulation results conducted on image transmission demonstrate that the proposed SNR-EQ-JSCC outperforms the state-of-the-art SwinJSCC in peak signal-to-noise ratio (PSNR) and perception metrics \textcolor{black}{while only requiring 0.05\% of the storage overhead and 6.38\% of the computational complexity for CA.} \textcolor{black}{Moreover, the channel-adaptive query method demonstrates significant improvements in perception metrics. When instantaneous SNR feedback is imperfect, SNR-EQ-JSCC using only the average SNR still surpasses baseline schemes. }


\end{abstract}



\IEEEpeerreviewmaketitle

\vspace{-0.4cm}
\section{Introduction} \label{sec:intro}
\vspace{-0.1cm}


\IEEEPARstart{S}{emantic} communication is a promising paradigm for future wireless communications. Leveraging neural network (NN)-enabled joint source-channel coding (JSCC) techniques, semantic communication can efficiently transmit the meaning of source data (i.e., semantic information). It has demonstrated significant potential to surpass conventional communications in both reliability and transmission efficiency. \par

As an early work on JSCC, a convolutional neural network (CNN)-based architecture called DeepJSCC is proposed in \cite{bourtsoulatze2019deep}, which has demonstrated superior image reconstruction performance over traditional digital communications, especially at low signal-to-noise ratios (SNRs). However, CNNs compress all elements indiscriminately, leading to a limited ability to extract semantic information. To address this issue, Transformer-based semantic coding methods have been proposed in \cite{SwinJSCC,xie2021task,9791398}. Unlike convolutional layers that can only capture local information due to their fixed-sized convolutional kernels, the attention mechanism in the Transformer allows the semantic encoder to focus on extracting multi-scale semantic information from source data. Consequently, Transformer-based semantic coding approaches consistently outperformed CNN-based methods. \par

To achieve JSCC, semantic communications need to cope with the impact of dynamic channels through channel adaptation (CA). \textcolor{black}{One approach is to adaptively adjust the encoder's output based on feedback from the channel output \cite{10500305}. However, the overhead associated with this feedback is excessively high. Other methods depend on feedback channel state information to achieve CA,} which can be categorized into three types. The first type embeds the SNR \textit{before} compression \cite{liang2023semantic,bian2023deepjscc}, though the SNR information may be lost during the lossy compression process. The second type refines feature vectors \textit{after} data compression \cite{SwinJSCC}. However, this kind of approach may lead to undesired information loss during the non-channel-adaptive compression process. The third type achieves CA \textit{during} the compression process through the channel attention mechanism \cite{xu2021wireless,zhang2023predictive,wu2022channel}. These methods enhance the utilization of the SNR information by embedding it in each channel attention layer. Most importantly, these existing methods achieve CA with unaffordable costs of storage overhead and computational complexity. \par


To this end, we propose a novel architecture called SNR-EQ-JSCC in this paper, which can be seamlessly applied to Transformer-based semantic coding methods. \textcolor{black}{Our contribution lies in designing an efficient method for leveraging SNR information through a lightweight NN structure that achieves channel adaptability. The innovations of the proposed method are two-fold}. First, a channel-adaptive query (CAQ) method is proposed, which \textcolor{black}{explicitly} adjusts attention scores in the multi-head attention (MHA) block according to the SNR \textcolor{black}{information}. Second, inspired by \textcolor{black}{positional embedding} in \textcolor{black}{Transformers} \cite{vaswani2017attention}, we propose \textcolor{black}{an SNR embedding} method, which embeds the SNR into the inputs of the MHA block to enhance CA. To \textcolor{black}{further} stabilize the training process, penalty terms corresponding to the CAQ \textcolor{black}{method} are introduced. \textcolor{black}{Besides, considering that instantaneous SNR feedback may be imperfect, we propose an alternative method that uses only the average SNR, substituting it for the instantaneous SNR in SNR-EQ-JSCC without the need for retraining.} Simulation results conducted on image transmission \textcolor{black}{show} that our proposed SNR-EQ-JSCC outperforms SwinJSCC \cite{SwinJSCC} across a wide range of channel conditions, \textcolor{black}{with storage overhead and computational complexity for CA in SNR-EQ-JSCC being 0.05\% and 6.38\% of those in the CA module of SwinJSCC (Channel ModNet), respectively.} In addition, the CAQ method significantly enhances perception metrics. When instantaneous SNR feedback is imperfect, SNR-EQ-JSCC using only the average SNR outperforms SwinJSCC and ADJSCC \cite{xu2021wireless}. \par


\begin{figure*}[tb]
\centering
\subfigure[The architecture of the SNR-EQ-JSCC.]{
\includegraphics[scale=0.36]{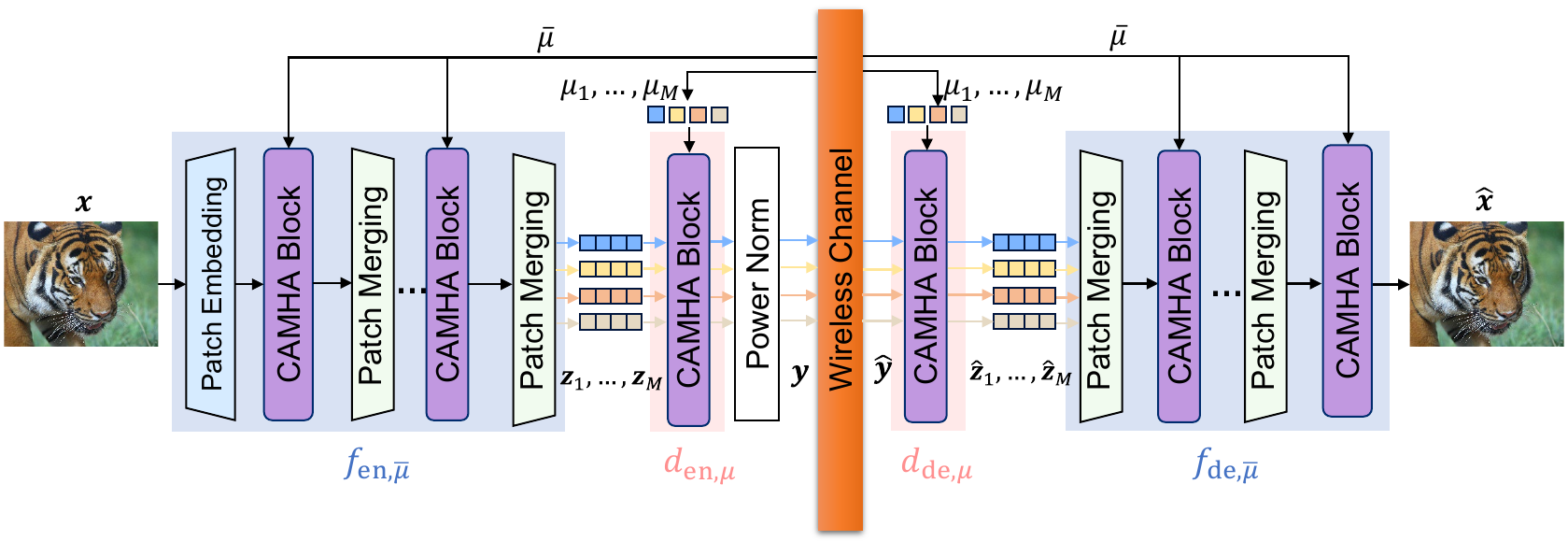} \vspace{-1cm} \label{fig:system1}  }
\subfigure[The structure of the CAMHA block.]{
\includegraphics[scale=0.385]{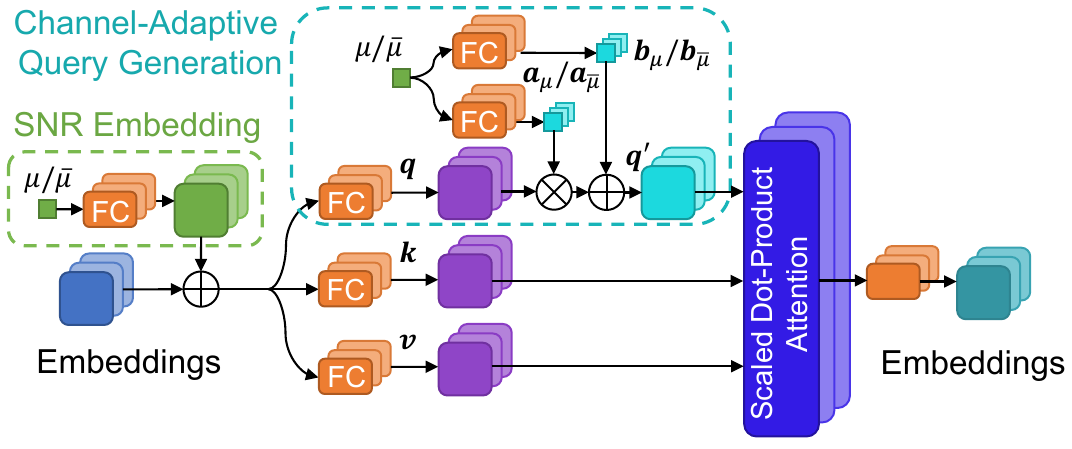} \label{fig:system2}}
\vspace{-0.3cm}
\caption{Illustration of the proposed SNR-EQ-JSCC.}
\vspace{-0.7cm}
\end{figure*}

\vspace{-0.3cm}
\section{System Model} \label{sec:system}  
\vspace{-0.1cm}

In this paper, we consider an image semantic coding system with SNR feedback, as illustrated in Fig. \ref{fig:system1}. In this system, the average SNR $\overline{\mu} \in \mathbb{R}$ \textcolor{black}{is} estimated at the receiver without error and then fed back to the transmitter. \textcolor{black}{Then, to enable the sender to obtain the instantaneous SNR $\mu$ given $\overline{\mu}$, the receiver feedback the fading coefficient.} \textcolor{black}{During the transmission of one image, the instantaneous SNR changes $M$ times while the average SNR stays the same.}   \par


The source \textcolor{black}{image is} denoted as $\boldsymbol{x} \in \mathbb{R}^{3 \times U \times W}$, where $U$ and $W$ denote \textcolor{black}{its} height and width, respectively. The coded \textcolor{black}{results are} denoted as $\boldsymbol{y} \in \mathbb{R}^{d_{\boldsymbol{y}}}$, where $ d_{\boldsymbol{y}}$ is the dimension of $\boldsymbol{y}$ and also the number of channel uses. For the noisy channel, the channel noise is denoted as $\boldsymbol{n} \in \mathbb{R}^{d_{\boldsymbol{y}}}$. Each component of $\boldsymbol{n}$ is \textcolor{black}{sampled} independently from a Gaussian distribution, i.e., $\boldsymbol{n} \sim \mathcal{N}\left(\boldsymbol{0}, \sigma_{\boldsymbol{n}}^2 \mathbf{I}_{d_{\boldsymbol{y}} \times d_{\boldsymbol{y}}}\right)$, where $\mathbf{I}_{d_{\boldsymbol{y}}\times d_{\boldsymbol{y}}}$ denotes the identity matrix \textcolor{black}{of size} $d_{\boldsymbol{y}}\times d_{\boldsymbol{y}}$ and $\sigma_{\boldsymbol{n}}^2$ denotes the noise power. The channel fading coefficient within the \textcolor{black}{$j$-th fading block is denoted as $h_j$, where $j\in\{1,2,\cdots,M\}$. Additionally, it is assumed that the expected value of $\mathbb{E}\left(h^2\right) = 1$, which implies that $\mu = \overline{\mu} \times h_j^2$.} \par

\textcolor{black}{Due to the fast fading, the entire image cannot be encoded based on instantaneous SNR. Therefore, we encode the entire image using the average SNR to leverage global information, while also further processing corresponding partial data based on the instantaneous SNR to address the impact of instantaneous fading. Specifically}, the transmitter first encodes the source image into \textcolor{black}{intermediate variables $\boldsymbol{z}$ according to $\overline{\mu}$, which is represented as follows}. 
\vspace{-0.4cm}

\begin{equation}
    \textcolor{black}{\boldsymbol{z} = f_{{\rm en},\overline{\mu}}(\boldsymbol{x},\overline{\mu}),} 
    \vspace{-0.2cm}
\end{equation}
where \textcolor{black}{$\boldsymbol{z} = \{\boldsymbol{z}_1,\cdots,\boldsymbol{z}_M\} $. Then, each $\boldsymbol{z}_j$ is further encoded into $\boldsymbol{y}_j$ based on the corresponding $\mu_j$, i.e., $\boldsymbol{y}_j = d_{{\rm en},\mu}(\boldsymbol{z}_j,\mu_j)$.} To satisfy the average power constraint of the transmitted signal, \textcolor{black}{each $\boldsymbol{y}_j$} is scaled such that \textcolor{black}{$\frac{M}{d_{\boldsymbol{y}}}\mathbb{E}(\boldsymbol{y}_j\boldsymbol{y}_j^T)\leq 1$} and then sent to the channel directly to ensure its transmission spans within a coherent time interval, and the received signal can be represented as follows,
\vspace{-0.4cm}

\begin{equation}
    \textcolor{black}{\boldsymbol{\hat{y}}_j = h_j \boldsymbol{y}_j + \boldsymbol{n}}.
    \vspace{-0.2cm}
\end{equation} \par

After receiving \textcolor{black}{$\boldsymbol{\hat{y}}_j$}, the receiver \textcolor{black}{first decodes $\boldsymbol{\hat{y}}_j$ according to $\mu_j$ to get $\boldsymbol{\hat{z}}_j$, i.e., $\boldsymbol{\hat{z}}_j = d_{{\rm de},\mu}(\boldsymbol{y}_j,\mu_j)$. Once the receiver recovers the complete $\boldsymbol{\hat{z}}$, it reconstructs the source image, which} can be represented as follows,
\vspace{-0.4cm}

\begin{equation}
    \textcolor{black}{\boldsymbol{\hat{x}} = f_{{\rm de},\overline{\mu}}(\boldsymbol{\hat{z}},\overline{\mu}).}
    \vspace{-0.2cm}
\end{equation}


When $ \mu $ feedback is imperfect, we substitute $\overline{\mu} $ for $ \mu $ in $d_{{\rm en},\mu}$ and $d_{{\rm de},\mu}$, which is treated as an alternative method for the proposed method without requiring retraining the NN.

\vspace{-0.3cm}
\section{SNR-EQ-JSCC} \label{sec:SNR-EQ-JSCC}
\vspace{-0.1cm}

Different from traditional Transformer-based semantic coding methods, SNR-EQ-JSCC leverages channel-adaptive MHA (CAMHA) blocks, as illustrated in Fig. \ref{fig:system2}. This block incorporates two main innovations: \textcolor{black}{the} CAQ and \textcolor{black}{SNR embedding methods}, which are elaborated in the following. \textcolor{black}{For simplicity, we take the CAMHA block based on the average SNR as an example.} \par


\vspace{-0.2cm}
\subsection{Preliminary of Multi-Head Attention Mechanism}

We first briefly \textcolor{black}{introduce} the scaled dot-product attention block and the \textcolor{black}{MHA} block \cite{vaswani2017attention}. \textcolor{black}{In the scaled dot-product attention block, input variables} are first mapped into queries $\boldsymbol{q} \in \mathbb{R}^{m_{\boldsymbol{q}}\times d_{\boldsymbol{q}}}$, keys $\boldsymbol{k} \in \mathbb{R}^{m_{\boldsymbol{q}}\times d_{\boldsymbol{q}}}$, and values $\boldsymbol{v} \in \mathbb{R}^{m_{\boldsymbol{q}}\times d_{\boldsymbol{v}}}$ through matrix multiplication, respectively. Here, $ d_{\boldsymbol{q}}$ and $ d_{\boldsymbol{v}}$ \textcolor{black}{denote} the dimensions of $\boldsymbol{q}/\boldsymbol{k}$ and $\boldsymbol{v}$, $ m_{\boldsymbol{q}}$ denotes the number of queries, keys, and values. Then, the outputs of the scaled dot-product attention block are computed by
\vspace{-0.3cm}

\begin{equation}
    {\rm Atten}\left(\boldsymbol{q}, \boldsymbol{k}, \boldsymbol{v}\right) = {\rm softmax}\left( \frac{\boldsymbol{q}\boldsymbol{k}^T}{\sqrt{d_{\boldsymbol{q}}}} \right)\boldsymbol{v}, \label{eq:attention}
\end{equation}
where ${\rm softmax}\left(\cdot \right)_{i,j} \triangleq \frac{{\rm e}^{{\left(\cdot \right)_{i,j}}}}{\sum_{i} \sum_{j} {\rm e}^{{\left(\cdot \right)_{i,j}}}  } $. Additionally, the \textit{attention score} is defined as ${\rm softmax}\left( \frac{\boldsymbol{q}\boldsymbol{k}^T}{\sqrt{d_{\boldsymbol{q}}}} \right)$, which indicates the importance of elements. \par

In the Transformer, the MHA block, an extension of the scaled dot-product attention block, is commonly applied. \textcolor{black}{For an} MHA block with $m_{\rm he}$ heads, \textcolor{black}{the $j$-th attention head applies separate linear transformations to the queries, keys, and values}. This process can be represented as follows,
\vspace{-0.45cm}

\begin{equation}
    \boldsymbol{q}^j \! = \boldsymbol{q}\boldsymbol{w}_{\boldsymbol{q}}^j, \boldsymbol{k}^j \! = \boldsymbol{k}\boldsymbol{w}_{\boldsymbol{k}}^j, \boldsymbol{v}^j \! = \boldsymbol{v}\boldsymbol{w}_{\boldsymbol{v}}^j  (j \in \{1,2, \cdots, m_{\rm he} \}),
\end{equation}
where $\boldsymbol{w}_{\boldsymbol{q}}^j \in \mathbb{R}^{ d_{\rm out}\times d_{\boldsymbol{q}}}$ and $\boldsymbol{w}_{\boldsymbol{k}}^j \in \mathbb{R}^{ d_{\rm out}\times d_{\boldsymbol{q}}}$, and $\boldsymbol{w}_{\boldsymbol{v}}^j \in \mathbb{R}^{ d_{\rm out}\times d_{\boldsymbol{v}}}$, with $d_{\rm out}$ denoting the output dimensions of the MHA block. Then, the attention score of the $i$-th head ${\rm Atten}_i$ can be calculated by Eq. \eqref{eq:attention}. Finally, a linear transformation is applied to the concatenation of all attention values to generate the outputs of the MHA block as
\vspace{-0.5cm}

\begin{equation}
    {\rm MHA}(\boldsymbol{q}, \boldsymbol{k}, \boldsymbol{v}) = {\rm concat}\left({\rm Atten}_1, \cdots , {\rm Atten}_{m_{\rm he}} \right)\! \boldsymbol{w}^{\rm out},
\end{equation}
where $\boldsymbol{w}^{\rm out} \! \in \! \mathbb{R}^{ \left(m_{\rm he} d_{\boldsymbol{q}}\right)\times d_{\rm out}}$ is the weight matrix of the outputs.

\vspace{-0.4cm}
\subsection{Channel-Adaptive Query}

The traditional MHA block generates the attention scores based solely on the source data. In semantic communications where the channel state varies dynamically, the MHA block should adjust attention scores according to the SNR. As mentioned above, the key intermediate variables of the MHA block \textcolor{black}{are} $\boldsymbol{q}$, $\boldsymbol{k}$, and $\boldsymbol{v}$. Since attention scores are independent of $\boldsymbol{v}$, the channel-adaptive $\boldsymbol{v}$ is not utilized. Additionally, due to the multiplicative relationship between $\boldsymbol{q}$ and $\boldsymbol{k}$, the impact of the SNR on $\boldsymbol{k}$ can be incorporated into its effect on $\boldsymbol{q}$. Hence, we \textcolor{black}{utilize} the \textcolor{black}{CAQ}, which are generated by
\vspace{-0.3cm}

\begin{equation}
    \boldsymbol{q}' = a_{\overline{\mu}} \boldsymbol{q} + b_{\overline{\mu}},
\end{equation}
where $a_{\overline{\mu}} = g_{\boldsymbol{\theta}_{\boldsymbol{q},1}}\left(\overline{\mu} \right) \in \mathbb{R}^+$ and $b_{\overline{\mu}} = g_{\boldsymbol{\theta}_{\boldsymbol{q},2}}\left(\overline{\mu} \right) \in \mathbb{R}$. Meanwhile, $\boldsymbol{\theta}_{\boldsymbol{q},1}$ and $\boldsymbol{\theta}_{\boldsymbol{q},2}$ are the parameters of the NNs that generate $a_{\overline{\mu}}$ and $b_{\overline{\mu}}$, respectively. The constraint $a_{\overline{\mu}} > 0$ is imposed to prevent the reversal of attention score rankings, which \textcolor{black}{leads to} instability during the training process. This constraint is \textcolor{black}{enforced} by applying the ReLU activation function at the output layer of $g_{\boldsymbol{\theta}_{\boldsymbol{q},1}}$. \par

\vspace{-0.3cm}
\subsection{\textcolor{black}{SNR Embedding}} \label{sec:search}
\vspace{-0.1cm}

Inspired by the \textcolor{black}{positional embedding} method in the Transformer, we incorporate the SNR into the inputs of the MHA block. Unlike the positional sequence, the SNR lacks sequential information. Therefore, the traditional \textcolor{black}{positional embedding} method based on the cosine function is unsuitable for \textcolor{black}{SNR embedding}. Instead, we propose the \textcolor{black}{SNR embedding} method inspired by learnable \textcolor{black}{positional embedding} methods:
\vspace{-0.3cm}

\begin{equation}
    \boldsymbol{x}_{\rm model}' = \boldsymbol{x}_{\rm model} + g_{\boldsymbol{\theta}_{\rm SNR}}(\overline{\mu}),
\end{equation}
where $\boldsymbol{x}_{\rm model}$ denotes the inputs of the \textcolor{black}{CAMHA} block and $g_{\boldsymbol{\theta}_{\rm SNR}}$ denotes the NN generating the SNR embeddings. By embedding the SNR into the inputs of the \textcolor{black}{CAMHA} block, the CA of SNR-EQ-JSCC can be further enhanced.

\vspace{-0.3cm}
\subsection{Loss Function of SNR-EQ-JSCC}
\vspace{-0.1cm}

Since \textcolor{black}{$a_{\overline{\mu}}$} and \textcolor{black}{$b_{\overline{\mu}}$} directly influence the attention scores in all \textcolor{black}{CAMHA} blocks, they significantly impact semantic coding performance. To stabilize the training process after incorporating SNR, penalty terms are introduced into the loss function. \par

At a lower \textcolor{black}{$\overline{\mu}$}, SNR-EQ-JSCC needs to omit more non-essential elements to provide more protection for the crucial elements containing semantic information. To this end, as $\mu$ decreases, SNR-EQ-JSCC needs to accomplish two objectives. First, it needs to increase the differentiation in attention scores, which requires increasing \textcolor{black}{$a_{\overline{\mu}}$}. Second, it needs to increase the proportion of elements with low attention scores, which requires increasing \textcolor{black}{$b_{\overline{\mu}}$}. Correspondingly, it needs to ensure that both \textcolor{black}{$a_{\overline{\mu}}$} and \textcolor{black}{$b_{\overline{\mu}}$} are negatively correlated with \textcolor{black}{$\overline{\mu}$}. Concerning \textcolor{black}{$a_{\overline{\mu}}$}, the following penalty term is introduced. \par
\vspace{-0.3cm}

\begin{equation}
    L_a\left(\overline{\mu}, a_{\overline{\mu}}\right) = {\rm ReLU}\left( {\rm Corr} \left(\overline{\mu}, a_{\overline{\mu}}\right) \right),
\end{equation}
where ${\rm Corr} \left(\cdot \right)$ denotes the Pearson correlation function \textcolor{black}{and ${\rm ReLU}(\cdot) \triangleq \max(0,\cdot) $}. Directly using \textcolor{black}{correlation function} as a penalty term is unsuitable, as this would drive the correlation coefficient between \textcolor{black}{$\overline{\mu}$} and \textcolor{black}{$a_{\overline{\mu}}$} to -1, while it is sufficient for them to be negatively correlated. The step function is not used to avoid drastic gradient changes that could destabilize the training process. Following this, we define $L_b$ in the same manner, which can be represented as follows,
\vspace{-0.3cm}

\begin{equation}
    L_b\left(\overline{\mu}, b_{\overline{\mu}}\right) = {\rm ReLU}\left( {\rm Corr} \left(\overline{\mu}, b_{\overline{\mu}}\right) \right).
\end{equation} \par

Hence, the loss function of SNR-EQ-JSCC is expressed as
\vspace{-0.3cm}

\begin{align}
    L\left(\boldsymbol{x}, \hat{\boldsymbol{x}}, \overline{\mu}, a_{\overline{\mu}}, b_{\overline{\mu}} \right) = & MSE\left(\boldsymbol{x}, \hat{\boldsymbol{x}}\right) \nonumber \\
    & + \lambda \times \left( L_a\left(\overline{\mu}, a_{\overline{\mu}}\right) + L_b\left(\overline{\mu}, b_{\overline{\mu}}\right)  \right),  
\end{align} 
where $MSE\left(\boldsymbol{x}, \hat{\boldsymbol{x}}\right)$ aims to reconstruct the source image and $\lambda > 0$ is a large constant.





\section{Experiment Results} \label{sec:results}

\vspace{-0.15cm}
\subsection{Experiment Settings}
\vspace{-0.1cm}

The experiments are conducted on an Intel Xeon Silver 4310 CPU and eight 48 GB Nvidia RTX A40 graphics cards. We compare the following schemes in the experiments.\footnote{\textcolor{black}{For fair comparison, all methods train a single model rather than training models for each average testing SNR $\overline{\mu}_{\rm test}$.}}

\begin{itemize}
    \item \textit{SNR-EQ-JSCC}: This method employs both \textcolor{black}{SNR embedding} and CAQ. \textcolor{black}{The backbone follows the design of SwinJSCC \cite{SwinJSCC} without its CA module, but the attention block near the channel is modified to align with the divided $\boldsymbol{z}$.}

    \item \textcolor{black}{\textit{SNR-EQ-JSCC w/ $\overline{\mu}$}: This method is an alternative method of the \textit{SNR-EQ-JSCC}, which replaces $\mu$ with $\overline{\mu}$ in the inputs of the \textit{SNR-EQ-JSCC}.}

    \item \textit{SNR-EQ-JSCC w/o CAQ}: This method is the \textit{SNR-EQ-JSCC} without CAQ.
    
    \item \textit{SNR-EQ-JSCC w/o EM}: This method is the \textit{SNR-EQ-JSCC} without \textcolor{black}{SNR embedding}.

    \item \textcolor{black}{\textit{SNR-EQ-JSCC w/o CA}: This method is the \textit{SNR-EQ-JSCC} without CAQ and SNR embedding.}
    
    \item \textit{SwinJSCC} \cite{SwinJSCC}: This method utilizes a fully connected layer-based network to achieve CA.
    
    \item \textit{SwinJSCC w/o CA}: This method applies the SwinJSCC architecture without \textcolor{black}{its} CA \textcolor{black}{module}.
    
    \item \textit{ADJSCC} \cite{xu2021wireless}: This method is based on CNN, which achieves CA by the channel attention mechanism.
\end{itemize}


\begin{table*}[tb]
\centering
\caption{Parameters of the convolution layers in the \textit{ADJSCC} (output channels, kernel size, stride, padding).}
\vspace{-0.15cm}
\label{tab:parameter}
\scalebox{1}{%
\begin{tabular}{c|c c c c c c }
\hline
Encoder & (\textcolor{black}{128},8,2,2) & (\textcolor{black}{256},7,2,2) & (\textcolor{black}{512},7,2,2) & (\textcolor{black}{1024},7,2,2) & (\textcolor{black}{2048},\textcolor{black}{5},1,\textcolor{black}{0}) & ($3072\times r$,\textcolor{black}{4},1,\textcolor{black}{0}) \\ \hline
Decoder & (\textcolor{black}{2048},7,1,\textcolor{black}{0}) & (\textcolor{black}{1024},\textcolor{black}{5},2,\textcolor{black}{0}) & (\textcolor{black}{512},\textcolor{black}{3},2,\textcolor{black}{0}) & (\textcolor{black}{256},\textcolor{black}{3},2,\textcolor{black}{1}) & (\textcolor{black}{128},\textcolor{black}{4},2,\textcolor{black}{0}) & (3,\textcolor{black}{5},1,\textcolor{black}{0}) \\ \hline
\end{tabular} %
}
\vspace{-0.45cm}
\end{table*}

The experiments are conducted on the widely-used DIV2K dataset \cite{agustsson2016challenge}, which comprises 1,000 RGB images with about 2K resolution. Therein, the training, validation, and testing datasets consist of 800, 100, and 100 images, respectively. \textcolor{black}{To standardize the image sizes}, they are randomly cropped into patches with \textcolor{black}{resolutions} of $256\times 256$. In the experiments, we define the compression rate (CR) as
\vspace{-0.3cm}

\begin{equation}
    r \triangleq \frac{d_{\boldsymbol{y}}}{3\times U \times W}, \vspace{-0.1cm}
\end{equation}
\textcolor{black}{where $r=1/32$ and $r=1/8$ are set as typical examples of low and high compression rates, respectively.} We refine the structure of the ADJSCC model for the DIV2K dataset, whose detailed parameters are listed in Table \ref{tab:parameter}. In contrast, we retain the SwinJSCC structure, which was specifically designed for the DIV2K dataset. During training, we set $\overline{\mu}$ to be uniformly distributed within [-10,20] dB, and the channel fading magnitude $h$ follows a Rayleigh distribution. \textcolor{black}{Following 5G standards, we assume the transmission of an image spans $M=8$ fading blocks. Since the communication overhead for feedback is much lower than that for transmitting $\boldsymbol{y}$, the overhead for feedback is neglected in the experiments. } \par

We employ \textcolor{black}{three} metrics to evaluate the coding performance: peak SNR (PSNR), multi-scale structural similarity index (MS-SSIM)\textcolor{black}{, and learned perceptual image patch similarity (LPIPS)}. PSNR is defined as the ratio of the square of the possible maximum gray value to $MSE\left(\hat{\boldsymbol{x}},\boldsymbol{x} \right)$. It measures the image reconstruction quality at the pixel level. MS-SSIM measures the reconstruction quality from the perspective of luminance, contrast, and structure information. \textcolor{black}{LPIPS is an NN-based perceptual similarity metric designed to better align with human visual perception.}

\vspace{-0.4cm}
\subsection{Effects of $\lambda$ in the Loss Function}
\begin{figure}[tb]
\centering
\subfigure[\textcolor{black}{$r = 1/32$.}]{
\includegraphics[scale=0.285]{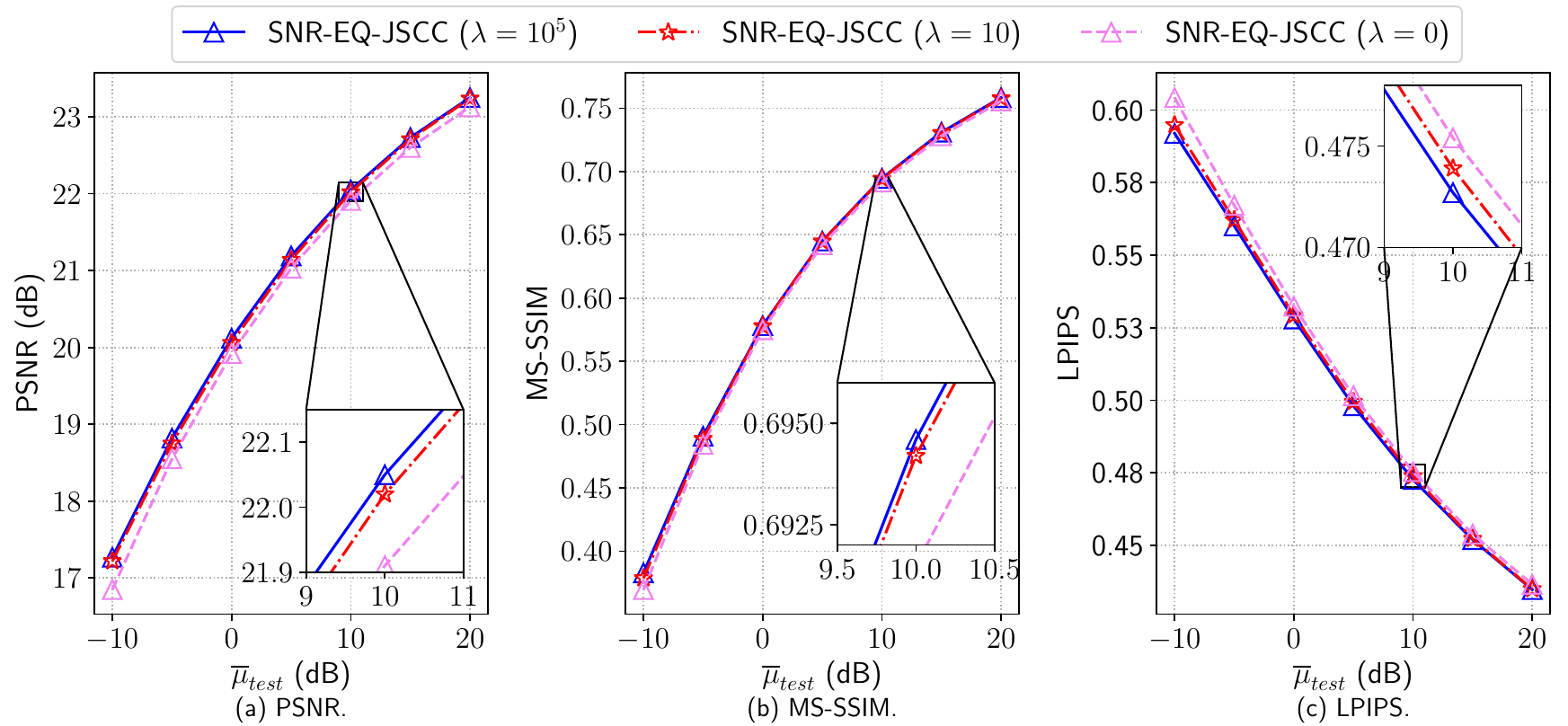} \vspace{-1cm} \label{fig:fig_psnr_lambda}  }
\subfigure[\textcolor{black}{$r = 1/8$.}]{
\includegraphics[scale=0.285]{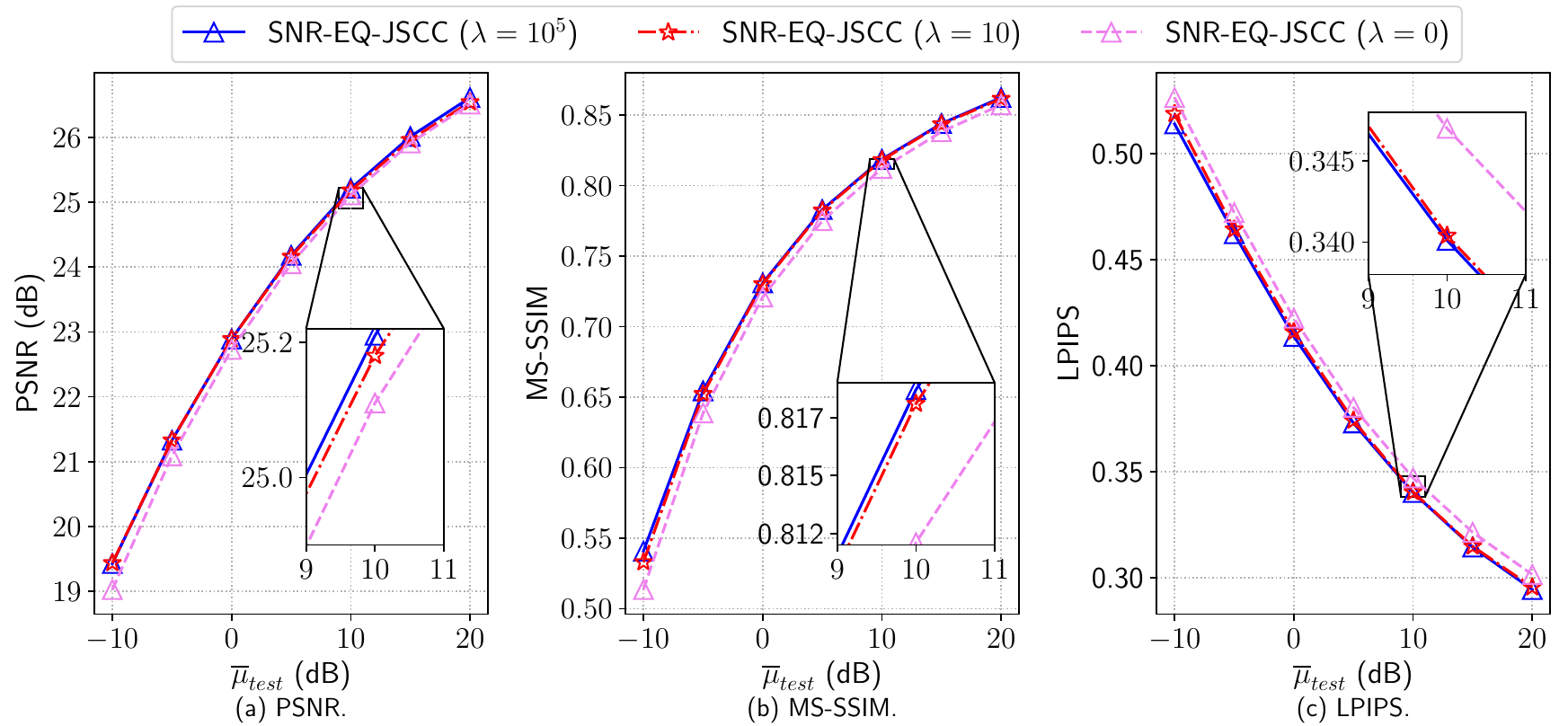} \label{fig:fig_ssim_lambda}}
\vspace{-0.3cm}
\caption{\textcolor{black}{Performance comparison of different $\lambda$.}}
\vspace{-0.7cm}
\end{figure}

\textcolor{black}{\textcolor{black}{Fig. \ref{fig:fig_psnr_lambda} and Fig. \ref{fig:fig_ssim_lambda} show performance of the \textit{SNR-EQ-JSCC} with different $\lambda$ at $r=1/32$ and $r=1/8$, respectively.} As illustrated, the decrease of $\lambda$ leads to a decline across all performance metrics. For example, when $r = 1/32 $ and $\overline{\mu}_{\rm test}=10$ dB, setting $\lambda = 10^5$ yields gains of
0.14 dB, 0.004, and 0.003 across the three metrics compared to $\lambda = 0$. When the penalty weight is zero, applying the CAQ method degrades subjective metrics such as MS-SSIM and LPIPS. This is because CAQ explicitly adjusts attention scores, which represent human focus. Therefore, unstable training can degrade perception metrics. In summary, a sufficiently large penalty weight is required to stabilize the training process.} \par

\begin{figure}[tb]
\centering
\subfigure[\textcolor{black}{$r = 1/32$.}]{
\includegraphics[scale=0.285]{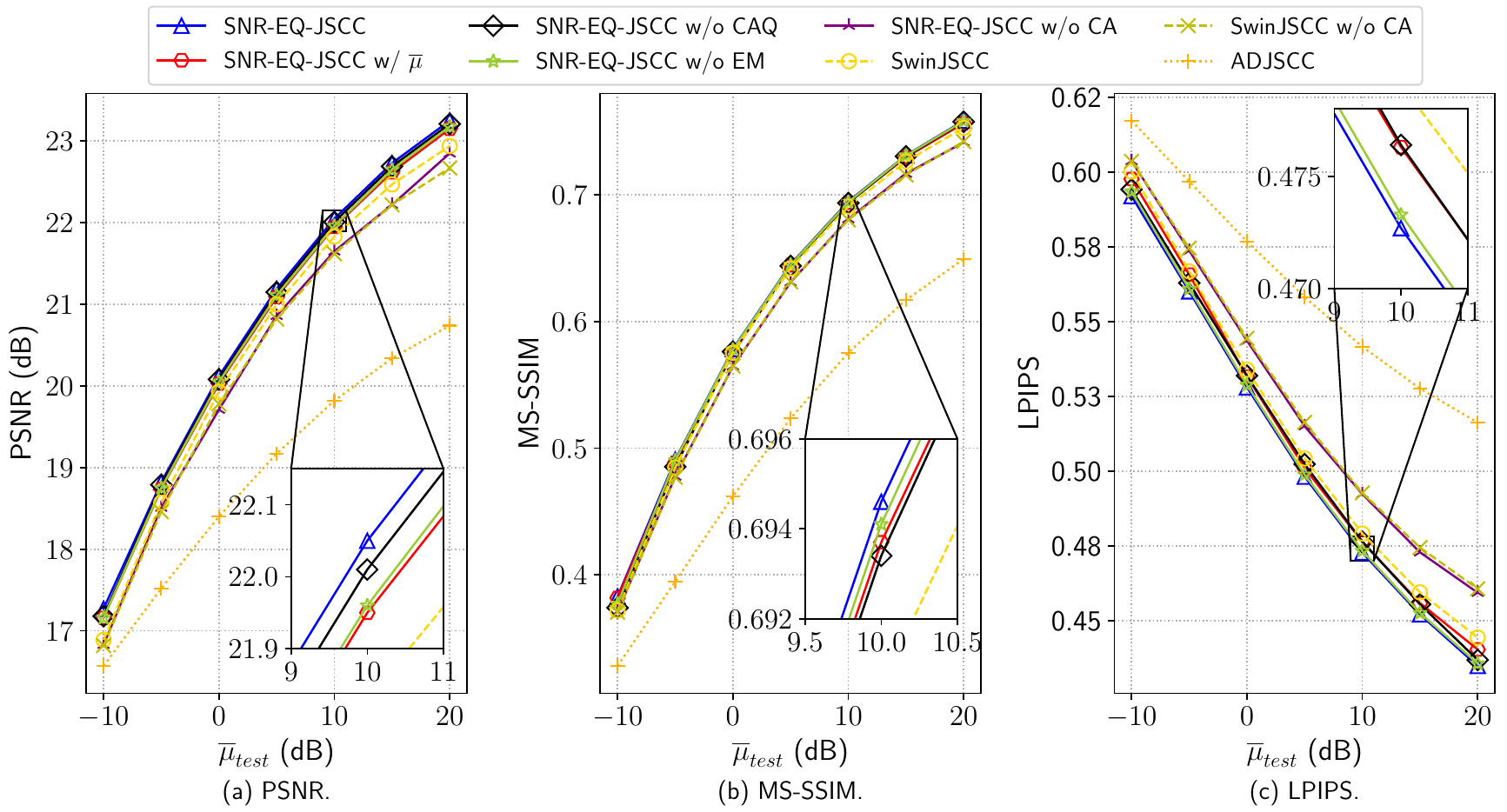} \label{fig:fig_psnr_div2k} \vspace{-0.3cm} }
\subfigure[\textcolor{black}{$r = 1/8$.}]{
\includegraphics[scale=0.285]{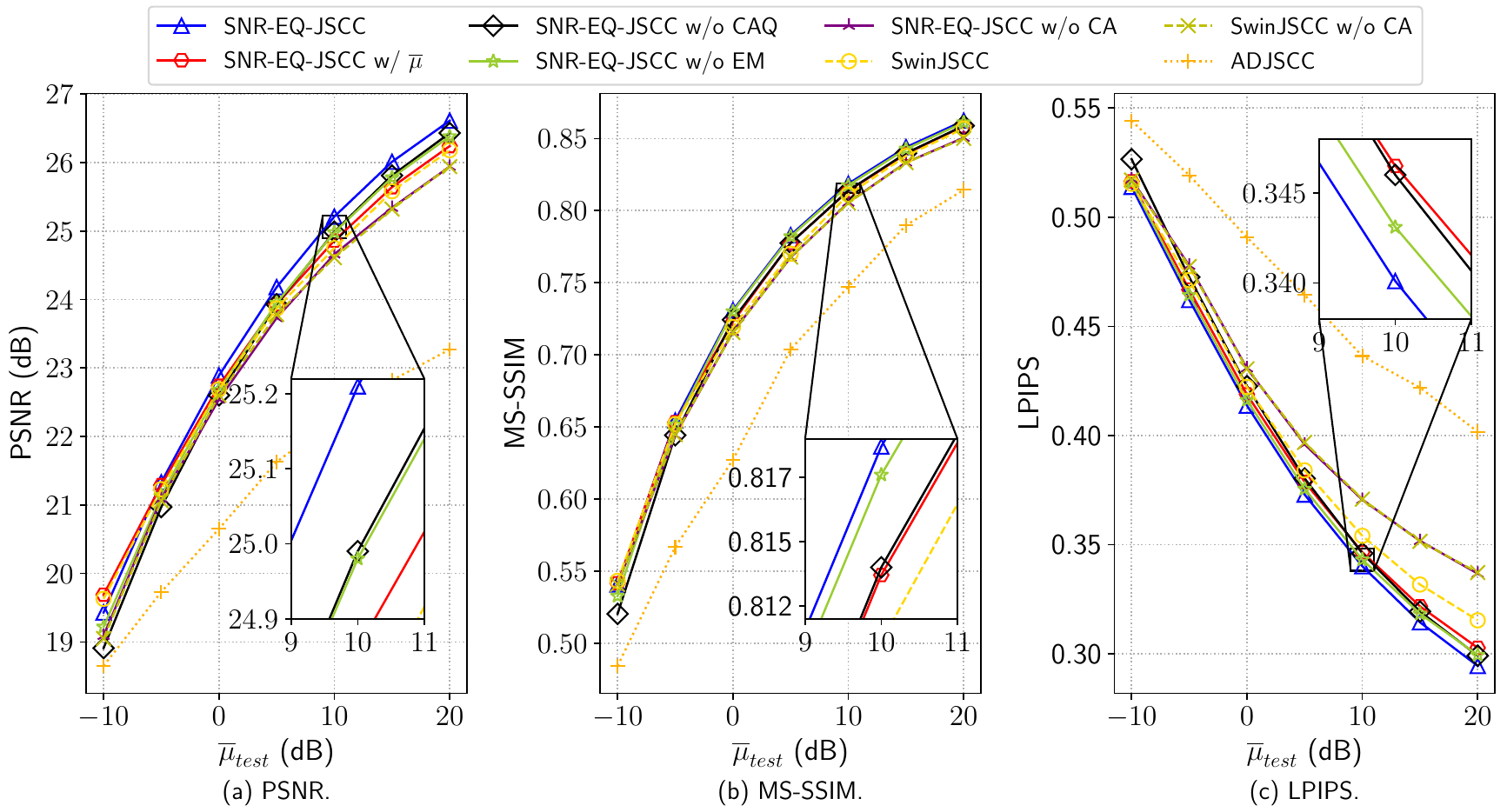} \label{fig:fig_ssim_div2k}}
\vspace{-0.3cm}
\caption{\textcolor{black}{Performance comparison at different $\overline{\mu}_{\rm test}$.}}
\vspace{-0.35cm}
\end{figure}

\vspace{-0.4cm}
\subsection{Performance Comparison of Different JSCC Schemes}

\textcolor{black}{Fig. \ref{fig:fig_psnr_div2k} and Fig. \ref{fig:fig_ssim_div2k} show performance at $r=1/32$ and $r=1/8$, respectively. Therein,} the \textit{SNR-EQ-JSCC} demonstrates significant performance gains compared to the \textit{SwinJSCC}. For example, at \textcolor{black}{$\overline{\mu}=20$} dB and \textcolor{black}{$r=1/32$}, the \textit{SNR-EQ-JSCC} outperforms the \textit{SwinJSCC} by \textcolor{black}{0.32} dB in PSNR, \textcolor{black}{0.0056} in MS-SSIM\textcolor{black}{, and 0.0095 in LPIPS}. This is because the SNR is incorporated into each attention block of the \textit{SNR-EQ-JSCC}. In contrast, the \textit{SwinJSCC} only utilizes the SNR to adjust the outputs of the last attention block after compression. Additionally, both the \textit{SNR-EQ-JSCC w/o CAQ} and the \textit{SNR-EQ-JSCC w/o EM} outperform the \textit{SwinJSCC}. Meanwhile, the performance of the \textit{SNR-EQ-JSCC w/o CAQ} is slightly higher than that of the \textit{SNR-EQ-JSCC w/o EM}. \textcolor{black}{Among the methods based solely on $\overline{\mu}$, the \textit{SNR-EQ-JSCC w/ $\overline{\mu}$} consistently outperforms other methods. For example, when $ r = 1/8 $ and $\overline{\mu}_{\rm test}=20$ dB, the \textit{SNR-EQ-JSCC w/ $\overline{\mu}$} achieves gains over the \textit{SwinJSCC} of 0.56 dB in PSNR, 0.009 in MS-SSIM, and 0.035 in LPIPS. }  \par

\textcolor{black}{Although the \textit{SNR-EQ-JSCC w/o EM} shows slightly lower PSNR performance than the \textit{SNR-EQ-JSCC w/o CAQ}, it achieves superior MS-SSIM and LPIPS. This is because the CAQ method explicitly adjusts attention scores, effectively using the attention mechanism to mimic human focus, thereby enhancing subjective metrics like MS-SSIM and LPIPS.}



\begin{figure*}[tb]
\centering
\includegraphics[scale=0.32]{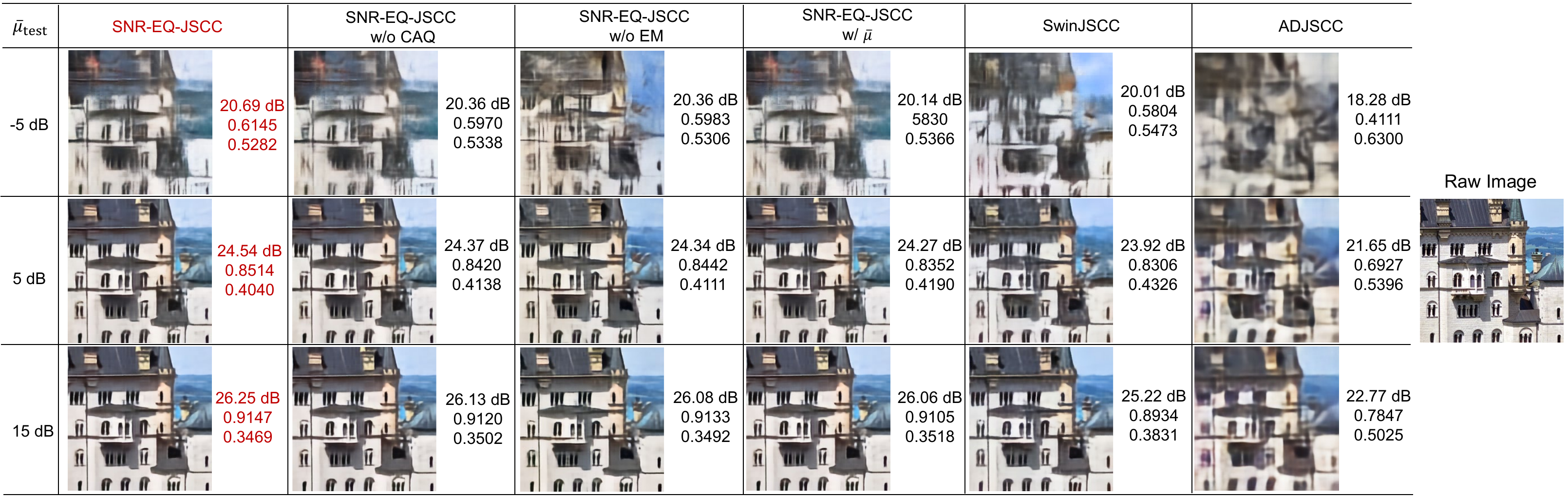}
\vspace{-0.2cm}
\caption{Illustration of the reconstructed images at \textcolor{black}{$r=1/32$}. The corresponding PSNR, MS-SSIM, and LPIPS results are presented beside the recovered image, respectively.}
\label{fig:recovered}
\vspace{-0.2cm}
\end{figure*}

\vspace{-0.4cm}
\subsection{Performance with Imperfect SNR Feedback}
\vspace{-0.05cm}

\begin{table}[tb]
\centering
\caption{\textcolor{black}{Performance with imperfect SNR feedback at $r=1/32$ and $\overline{\mu}_{\rm test}=0$ dB}.}
\vspace{-0.15cm}
\label{tab:performance}
\scalebox{0.9}{%
\begin{tabular}{c|c c c c c}
\hline
 & \textcolor{black}{$\sigma_h=0$}      & \textcolor{black}{$\sigma_h=0.1$}    &  \textcolor{black}{$\sigma_h=0.2$}    & \textcolor{black}{$\sigma_h=0.3$}     & \textcolor{black}{w/ $\overline{\mu}$}      \\ \hline
\textcolor{black}{PSNR (dB)}  & \textcolor{black}{20.18}  & \textcolor{black}{20.08}  & \textcolor{black}{\textit{19.77}}  & \textcolor{black}{19.34}   & \textcolor{black}{20.03}  \\ \hline
\textcolor{black}{MS-SSIM}    & \textcolor{black}{0.5782} & \textcolor{black}{0.5750} & \textcolor{black}{\textit{0.5579}} & \textcolor{black}{0.5332}  & \textcolor{black}{0.5746} \\ \hline
\textcolor{black}{LPIPS}      & \textcolor{black}{0.5280} & \textcolor{black}{0.5294} & \textcolor{black}{\textit{0.5340}} & \textcolor{black}{0.5428}  & \textcolor{black}{0.5316} \\ \hline
\end{tabular} %
}
\vspace{-0.65cm}
\end{table}


To account for unreliable feedback links, SNR quantization, and outdated information, we consider the scenario of imperfect feedback \textcolor{black}{for $\mu$}. Specifically, the feedback error for the fading coefficient is assumed to be a complex Gaussian distribution with variance $\sigma_h^2$. Note that due to \textcolor{black}{the relatively long persistence of} $\overline{\mu}$, it is assumed to be accurate. Since methods using only $\overline{\mu}$ are unaffected by imperfect SNR feedback, simulations for these methods are not repeated. For $r=1/32$ and $\overline{\mu}_{\rm test}=0$ dB, the results in Table \ref{tab:performance} indicate that when $\sigma_h$ exceeds 0.2, the performance of the \textit{SNR-EQ-JSCC} falls below that of the \textit{SNR-EQ-JSCC w/ $\overline{\mu}$}. In such cases, the \textit{SNR-EQ-JSCC w/ $\overline{\mu}$} needs to be applied, which only requires replacing $\mu$ with $\overline{\mu}$ in $d_{{\rm en},\mu}$ and $d_{{\rm de},\mu}$.



\vspace{-0.3cm}
\subsection{Visualization Results} \label{sec:vis}
\vspace{-0.05cm}

Fig. \ref{fig:recovered} presents the images \textcolor{black}{recovered by} various schemes at \textcolor{black}{$r=1/32$}. It illustrates that the \textit{SNR-EQ-JSCC} consistently maintains the highest clarity and the sharpest contour details, corresponding to the \textcolor{black}{superior} PSNR, MS-SSIM, \textcolor{black}{and LPIPS}. Meanwhile, it illustrates that the images reconstructed by the \textit{ADJSCC} at low SNR exhibit a blurring effect similar to that seen in low-resolution image interpolation \textcolor{black}{and noticeable block boundaries}. This is because CNNs can only extract and reconstruct based on limited local information.


\vspace{-0.4cm}
\subsection{Storage Overhead and Computational Complexity} \label{sec:complexity}
\vspace{-0.05cm}

We use the \textcolor{black}{additional} number of model parameters and that of floating point operations (FLOPs) to measure storage overhead and computational complexity \textcolor{black}{for achieving CA}, respectively. \textcolor{black}{The FLOPs and parameters at $r=1/32$ are listed in Table \ref{tab:div2k}, where the attention feature (AF) module is the CA module in ADJSCC. It demonstrates that the storage overhead and computational complexity of the CA module in the \textit{SNR-EQ-JSCC} are only 0.05\% and 6.38\% of those in the Channel ModeNet, respectively, and 17.58\% and 28.12\% of those in the AF module, respectively. }

\begin{table*}[t]
\centering
\caption{Storage overhead and computational complexity \textcolor{black}{at $r = 1/32$}.}
\vspace{-0.15cm}
\label{tab:div2k}
\scalebox{0.9}{%
\begin{tabular}{c|ccc|cc|cc}
\hline
\multirow{2}{*}{} & \multicolumn{3}{c|}{SNR-EQ-JSCC} & \multicolumn{2}{c|}{SwinJSCC} & \multicolumn{2}{c}{ADJSCC} \\ \cline{2-8} 
 & \multicolumn{1}{c|}{Backbone w/o CA} & \multicolumn{1}{c|}{SNR Embedding} & CAQ & \multicolumn{1}{c|}{Backbone w/o CA} & Channel ModNet & \multicolumn{1}{c|}{Backbone w/o CA \cite{kurka2020deepjscc}} & AF Module \\ \hline
FLOPs & \multicolumn{1}{c|}{$3.21\times 10^{10}$} & \multicolumn{1}{c|}{$3.93\times 10^5$} & $2.03\times 10^2$ & \multicolumn{1}{c|}{$3.27\times 10^{10}$} & $8.72\times 10^8$ & \multicolumn{1}{c|}{$1.49\times 10^{10}$} & $2.24\times 10^6$ \\ \hline
Parameters & \multicolumn{1}{c|}{$1.56\times 10^7$} & \multicolumn{1}{c|}{$6.29\times 10^5$} & $2.42\times 10^2$ & \multicolumn{1}{c|}{$1.83\times 10^7$} & $9.86\times 10^6$ & \multicolumn{1}{c|}{$1.61\times 10^7$} & $2.24\times 10^6$ \\ \hline
\end{tabular}
}
\vspace{-0.65cm}
\end{table*}

\vspace{-0.3cm}
\section{Conclusion} \label{sec:conclusion}
\vspace{-0.1cm}

In this paper, we propose a novel channel-adaptive semantic coding architecture called SNR-EQ-JSCC, which can be seamlessly applied to Transformer-based approaches. The main novelty \textcolor{black}{of this design} is two-fold. Firstly, SNR-EQ-JSCC adaptively adjusts query values according to the SNR. Secondly, SNR-EQ-JSCC embeds the SNR into the inputs of the attention block to enhance CA. The penalty terms are introduced to stabilize the training process. The simulations on the DIV2K datasets demonstrate that our proposed SNR-EQ-JSCC outperforms SwinJSCC, while the storage overhead and computational complexity for CA in the SNR-EQ-JSCC are only \textcolor{black}{0.05}\% and \textcolor{black}{6.38}\% of those in the Channel ModNet, respectively. \textcolor{black}{Furthermore, the CAQ method can significantly improve perception metrics}, while the storage overhead and computational complexity to generate CAQ are only \textcolor{black}{one hundred-thousandth and one ten-thousandth} of those in the Channel ModNet, respectively. In summary, this work sheds light on achieving CA with extremely low storage overhead and computational complexity. In future work, a potential direction is to extend SNR-EQ-JSCC for multi-carrier or multi-antenna scenarios.

\ifCLASSOPTIONcaptionsoff
  \newpage
\fi

\vspace{-0.1cm}
\bibliographystyle{IEEEtran}
\bibliography{refer.bib}



\end{document}